# On the Intrinsic Link between Gradient Strengthening and Passivation Onset in Single Crystal Plasticity

Habib Pouriayevali

Researcher, Germany,

this work was partially carried out during the DFG-funded project PO-2242/3-1 at Technische Universität Darmstadt.

**Abstract**

A finite-deformation framework for gradient crystal plasticity is developed within a thermodynamically consistent setting grounded in Gurtin's power-conjugate formulation. The model introduces a flow rule that accounts explicitly for both energetic and dissipative microstress contributions. Numerical simulations are performed to investigate the response of single crystals subjected to passivation-type boundary constraints. The results reveal that constitutive laws capable of reproducing size-dependent strengthening at the onset of plastic flow simultaneously generate a pronounced, nearly elastic-type response when passivation is imposed. These findings establish a fundamental connection between gradient-induced yield strengthening and boundary-driven elevation of the mechanical response, highlighting the essential influence of dissipative gradient effects.

# 1  Introduction

Classical crystal plasticity theories do not inherently account for experimentally observed size effects in crystalline materials. The need to incorporate length-scale sensitivity has therefore motivated the development of strain-gradient and gradient crystal plasticity formulations. One of the earliest contributions in this direction was provided by Aifantis (1984), and the subject has since evolved through significant theoretical advances (Fleck and Hutchinson, 1993; Huang et al., 2000; Forest, 2008; Kuroda and Tvergaard, 2008; Reddy et al., 2008; Gurtin and Anand, 2009). Numerous frameworks have been proposed to capture scale-dependent phenomena, and extensive efforts have been devoted to clarifying the mechanical role of gradient terms. Among these approaches, thermodynamically consistent formulations have attracted particular attention. A comprehensive overview of modern gradient plasticity theories is given by Voyiadjis and Song (2019).

Within gradient theories, microstresses associated with geometrically necessary dislocations (GNDs) generate two distinct mechanical effects: gradient strengthening and gradient hardening. The former manifests as an apparent extension of the elastic regime or a pronounced stiffness increase at the initiation of plastic flow. The latter corresponds to a progressive evolution of flow resistance driven by the accumulation of GNDs during continued deformation.

Another topic of continuing interest concerns the response of single crystals under passivation-type boundary conditions. Previous numerical investigations, notably those summarized by Evans and Hutchinson (2009), demonstrated that the presence of a passivated surface layer can significantly modify the initial yielding behavior. In particular, gradient-induced yield strengthening was interpreted as an effectively extended elastic response, with higher-order boundary tractions playing a central role in governing this behavior.

The present study examines whether the scale-dependent strengthening observed at the onset of plastic deformation is intrinsically linked to the elevated response that emerges when passivation is imposed during loading. Specifically, we address the question of whether a constitutive formulation capable of reproducing gradient strengthening at initial yield necessarily predicts the sharp increase in stiffness associated with the activation of passivation.

The remainder of the paper is organized as follows. Section 2 outlines the finite-deformation gradient crystal plasticity formulation in the reference configuration. Section 3 presents the numerical results, and Section 4 summarizes the main findings and conclusions.

## 2 Formulation

The formulation adopted in this work follows the finite-deformation gradient crystal plasticity framework introduced by Gurtin (2006, 2008, 2010). The theory is presented here in terms of the reference configuration. Although the development is carried out in a finite-deformation setting, the model may be reduced to the small-deformation regime when appropriate.

### 2.1 Kinematics in a single crystal

As a starting point,

$$\mathbf{F} = \mathbf{F}^e\mathbf{F}^p, \ \det \mathbf{F}^p = 1, \ \mathbf{C}^e = \mathbf{F}^{eT}\mathbf{F}^e, \ \mathbf{E}^e = \frac{1}{2}(\mathbf{C}^e - \mathbf{I})$$

$$\text{and} \ \mathbf{L} = \mathbf{L}^e + \mathbf{F}^e\mathbf{L}^p\mathbf{F}^{e-1} \tag{1}$$

are defined, where $\mathbf{F}$ represents the deformation gradient tensor, $\mathbf{F}^e$ and $\mathbf{F}^p$ denote the elastic and inelastic distortions, respectively. det stands for the determinant of a tensor. $\mathbf{C}^e$ is the right Cauchy-Green elastic tensor, $\mathbf{E}^e$ is the Green-Lagrangian elastic-strain tensor, and $\mathbf{I}$ is the identity tensor. $\mathbf{L}$ is the velocity gradient, and $\mathbf{L}^e$ and $\mathbf{L}^p$ represent the elastic and plastic distortion-rate tensors given by

$$\mathbf{L}^{\mathrm{e}} = \dot{\mathbf{F}}^{\mathrm{e}}\mathbf{F}^{\mathrm{e}-1}, \mathbf{L}^{\mathrm{p}} = \sum_{\alpha} v^{\alpha}\, \mathbb{S}^{\alpha}. \tag{2}$$

Here, $v^{\alpha}$ is a flow rate and $\mathbb{S}^{\alpha}$ is the Schmid tensor defined as

$$\mathbb{S}^{\alpha} = \mathbf{s}^{\alpha} \otimes \mathbf{m}^{\alpha},\ \mathrm{trace}\ \mathbb{S}^{\alpha} = 0\ \mathrm{and}\ \mathbf{l}^{\alpha} = \mathbf{m}^{\alpha} \times \mathbf{s}^{\alpha}, \tag{3}$$

where $\alpha$ is a slip-system number, $\mathbf{m}^{\alpha}$, $\mathbf{s}^{\alpha}$ and $\mathbf{l}^{\alpha}$ are unit vectors representing the normal vector of the slip plane and the glide direction of edge and screw dislocations, respectively. These vectors are constant in the intermediate space (Gurtin and Reddy, 2016) and vary in the reference and deformed configurations (Gurtin, 2006). In this study, $r$ stands for the reference configuration, and $\mathbf{s}_r^{\alpha}$, $\mathbf{m}_r^{\alpha}$ and $\mathbf{l}_r^{\alpha}$ are then given by

$$\mathbf{s}_r^{\alpha} = \mathbf{F}^{\mathrm{p}-1}\mathbf{s}^{\alpha},\ \mathbf{m}_r^{\alpha} = \mathbf{F}^{\mathrm{p}\mathrm{T}}\mathbf{m}^{\alpha}\ \mathrm{and}\ \mathbf{l}_r^{\alpha} = \mathbf{F}^{\mathrm{p}-1}\mathbf{l}^{\alpha}. \tag{4}$$

## 2.2 Internal and external powers, and force balances

Guided by the principle of virtual power (Gurtin, 2008, 2010), macroscopic and microscopic force balances, as well as macroscopic and microscopic traction conditions, are respectively derived as

$$\mathrm{Div}\,\mathbf{P} + \mathbf{b} = \mathbf{0},\ \tau^{\alpha} - \pi^{\alpha} + \mathrm{Div}\,\boldsymbol{\xi}^{\alpha} = 0, \tag{5}$$

$$\mathbf{p}(\mathbf{n}_r) = \mathbf{P}\mathbf{n}_r\ \mathrm{and}\ \xi(\mathbf{n}_r)^{\alpha} = \boldsymbol{\xi}^{\alpha} \cdot \mathbf{n}_r. \tag{6}$$

Here, $\mathbf{P}$ represents the first Piola-Kirchhoff stress and $\tau^{\alpha} = \mathbf{C}^{\mathrm{e}}\mathbf{S}^{\mathrm{e}}:\mathbb{S}^{\alpha}$ is the resolved shear stress, where $\mathbf{S}^{\mathrm{e}}$ represents the second Piola elastic-stress.

## 2.3 Measurement of dislocation densities

In the finite-deformation crystal plasticity description, by considering $\nabla^{\#} v^{\alpha} = \mathbf{F}^{\mathrm{p}-\mathrm{T}}\nabla v^{\alpha}$, where $\nabla^{\#}$ denotes gradient operation in the intermediate (lattice) space

$$\dot{\rho}_{\vdash}^{\alpha} = -\mathbf{s}_r^{\alpha} \cdot \nabla v^{\alpha} = -\mathbf{s}^{\alpha} \cdot \nabla^{\#} v^{\alpha},$$

$$\dot{\rho}_{\odot}^{\alpha} = \mathbf{l}_r^{\alpha} \cdot \nabla v^{\alpha} = \mathbf{l}^{\alpha} \cdot \nabla^{\#} v^{\alpha} \tag{7}$$

are directional derivatives of the slip rate in the direction of glides $-\mathbf{s}^{\alpha}$ and $\mathbf{l}^{\alpha}$.

## 2.4 Free-energy imbalance

Guided by the second law of thermodynamics and the balance of powers (Gurtin, 2008, 2010), the local dissipation inequality may be rewritten as

$$\dot{\Psi} - \mathbf{S}^e : \dot{\mathbf{E}}^e - \sum_\alpha (\pi^\alpha \nu^\alpha + \boldsymbol{\xi}^\alpha \cdot \nabla \nu^\alpha) \leq 0, \tag{8}$$

where $\Psi$ denotes a stored free-energy per unit reference volume and admits the decomposition

$$\Psi = \Psi^e + \Psi^\rho, \tag{9}$$

where $\Psi^e$ is an elastic energy related to the stretching of the crystal lattice and $\Psi^\rho$ is a defect free-energy. $\Psi^\rho$ presents the defect energy contributed by GNDs only.

### 2.4.1 Elastic energy $\Psi^e$

$\Psi^e$ is given in the form of

$$\Psi^e = \frac{\mu}{2}(I_1 - 3 - \ln I_3) + \frac{\lambda}{2}\left(\ln I_3^{1/2}\right)^2,$$

$$\dot{\Psi}^e = \frac{\partial \Psi^e}{\partial \mathbf{E}^e} : \dot{\mathbf{E}}^e \tag{10}$$

where $I_1$ and $I_3$ are the first and third invariants of $\mathbf{C}^e$, $\mu$ is the shear modulus, and $\lambda$ is the Lame parameter (Bonet and Wood, 2008).

## 2.5 Constitutive model

Considering the rate of the free energies introduced in the previous sections,

$$\dot{\Psi} = \frac{\partial \Psi^e}{\partial \mathbf{E}^e} : \dot{\mathbf{E}}^e + \sum_\alpha \left(f^\alpha_\vdash \dot{\rho}^\alpha_\vdash + f^\alpha_\odot \dot{\rho}^\alpha_\odot\right), \tag{11}$$

the local dissipation inequality in Eq. 8 is rewritten in

$$\dot{\Psi} - \mathbf{S}^e : \dot{\mathbf{E}}^e - \sum_\alpha (\pi^\alpha \nu^\alpha + \boldsymbol{\xi}^\alpha \cdot \nabla \nu^\alpha)$$

$$= \left(\frac{\partial \Psi^e}{\partial \mathbf{E}^e} - \mathbf{S}^e\right) : \dot{\mathbf{E}}^e + \sum_\alpha \left(-f^\alpha_\vdash \mathbf{s}^\alpha_r + f^\alpha_\odot \mathbf{l}^\alpha_r\right) \cdot \nabla \nu^\alpha \tag{12}$$

$$-\sum_\alpha (\pi^\alpha v^\alpha + \boldsymbol{\xi}^\alpha \cdot \nabla v^\alpha) \leq 0.$$

Here, the recoverable and non-recoverable energetic microstresses, as well as the dissipative one, are introduced as

$$\boldsymbol{\xi}^\alpha = \boldsymbol{\xi}^\alpha_{engR} + \boldsymbol{\xi}^\alpha_{dis}. \tag{13}$$

By substitution of Eq. 13 into Eq. 12, the following expressions associated with Eqs. 10, are found,

$$\frac{\partial \Psi^e}{\partial \mathbf{E}^e} = \mathbf{S}^e,$$

$$\boldsymbol{\xi}^\alpha_{engR} \cdot \nabla v^\alpha = \left(-f^\alpha_{\ominus} \mathbf{s}^\alpha_r + f^\alpha_{\odot} \mathbf{l}^\alpha_r\right) \cdot \nabla v^\alpha \tag{14}$$

A reduced form of inequality is also given by

$$\sum_\alpha (\pi^\alpha v^\alpha + \boldsymbol{\xi}^\alpha_{dis} \cdot \nabla v^\alpha) \geq 0. \tag{15}$$

Eq. 14 shows that the energetic microstresses $\boldsymbol{\xi}^\alpha_{engR}$ are in-slip-plane vectors. Consequently, $\boldsymbol{\xi}^\alpha_{dis}$ is assumed to be tangential to the slip plane $\Pi^\alpha$ and power terms in Eqs. 14 and 15 are redefined in

$$\boldsymbol{\xi}^\alpha_{engR} \cdot \nabla v^\alpha = \boldsymbol{\xi}^\alpha_{engR} \cdot \nabla^\alpha v^\alpha,$$

$$\sum_\alpha (\pi^\alpha v^\alpha + \boldsymbol{\xi}^\alpha_{dis} \cdot \nabla^\alpha v^\alpha) \geq 0, \tag{16}$$

where $\nabla^\alpha v^\alpha = \nabla v^\alpha - (\bar{\mathbf{m}}^\alpha_r \cdot \nabla v^\alpha)\bar{\mathbf{m}}^\alpha_r$, and $\bar{\mathbf{m}}^\alpha_r = \frac{\mathbf{m}^\alpha_r}{|\mathbf{m}^\alpha_r|}$.

A strong form of reduced inequality in Eq. 16 is also considered by

$$\pi^\alpha v^\alpha \geq 0,$$

$$\boldsymbol{\xi}^\alpha_{dis} \cdot \nabla^\alpha v^\alpha \geq 0. \tag{17}$$

As we know, the conventional rate-dependence plasticity theories are accompanied by a dissipative scalar microforce $\pi^\alpha$, power-conjugated to the flow rate $v^\alpha$. This microforce is commonly defined through a power-law term and takes a non-incremental form, found in Eq. 18. In the same way, the dissipative higher-order stress $\xi^\alpha_{dis}$ is also expressed in a rate-like form, found in Eq. 20. Thus

$$\pi^\alpha = S^\alpha \left(\frac{|v^\alpha|}{v_0}\right)^m \frac{v^\alpha}{|v^\alpha|}, \quad \pi^\alpha v^\alpha = S^\alpha \left(\frac{|v^\alpha|}{v_0}\right)^m |v^\alpha| \geq 0, \tag{18}$$

$$S^\alpha(0) = S_y, \tag{19}$$

$$\xi^\alpha_{dis} = L_2 S_d \left(\frac{L_2|\nabla^\alpha v^\alpha|}{d_0}\right)^q \frac{\nabla^\alpha v^\alpha}{|\nabla^\alpha v^\alpha|}, \tag{20}$$

where $m$ and $q$ are rate-sensitivity parameters, $v_0$ and $d_0$ are constant positive-valued references, $L_2$ is a dissipative length-scale parameter. $S^\alpha$ is a positive-valued slip resistance, $S_y$ is an initial value for $S^\alpha$, and $S_d$ is a positive-valued parameter.

### 2.5.1 Flow rule

Considering Eq. 13, the microscopic force balance given in Eq. (5)$_2$ takes the following form,

$$\tau^\alpha + \text{Div}\xi^\alpha_{engR} = \pi^\alpha - \text{Div}\xi^\alpha_{dis}. \tag{21}$$

The microstresses and the microforce defined in Eqs. 14, 18, and 20 are as follows:

$$\pi^\alpha = S^\alpha \left(\frac{|v^\alpha|}{v_0}\right)^m \frac{v^\alpha}{|v^\alpha|},$$

$$\xi^\alpha_{dis} = S_d L_2^{q+1} \left(\frac{|\nabla^\alpha v^\alpha|}{d_0}\right)^q \frac{\nabla^\alpha v^\alpha}{|\nabla^\alpha v^\alpha|} = S_d L_2^{q+1} \left(\frac{|\nabla^\alpha v^\alpha|}{d_0}\right)^q \frac{1}{|\nabla^\alpha v^\alpha|} \left(\frac{-\dot{\rho}^\alpha_\vdash s^\alpha_r}{|s^\alpha_r|^2} + \frac{\dot{\rho}^\alpha_\odot l^\alpha_r}{|l^\alpha_r|^2}\right),$$

$$\xi^\alpha_{engR} = S_0 L_1^2 \left(-\rho^\alpha_\vdash s^\alpha_r + \rho^\alpha_\odot l^\alpha_r\right). \tag{22}$$

Substitution of Eq. 22 into Eq. 21 results in

$$\tau^\alpha + \underbrace{S_0 L_1^2 \text{Div}\left(\left(-\rho^\alpha_\vdash s^\alpha_r + \rho^\alpha_\odot l^\alpha_r\right)\right)}_{(I)}$$

$$= \underbrace{S^\alpha \left(\frac{|v^\alpha|}{v_0}\right)^m \frac{v^\alpha}{|v^\alpha|}}_{(II)} - \underbrace{S_d L_2^{q+1} \text{Div}\left(\left(\frac{|\nabla^\alpha v^\alpha|}{d_0}\right)^q \frac{\nabla^\alpha v^\alpha}{|\nabla^\alpha v^\alpha|}\right)}_{(III)} \tag{23}$$

## 2.6 Implementation approach

In this study, the implementation of a two-dimensional version of the framework, where $\nabla v^\alpha$ and $\mathbf{l}^\alpha$ are respectively in- and out-of-plane vectors, is targeted. Thus $\dot{\rho}_\odot^\alpha = \mathbf{l}_r^\alpha \cdot \nabla v^\alpha = \mathbf{l}^\alpha \cdot \nabla^\# v^\alpha = 0$.

The model is implemented through a custom user element (UEL) developed in ABAQUS. In this approach, the evolution rate of the edge dislocation density, $\dot{\rho}_\vdash^\alpha = -\mathbf{s}_r^\alpha \cdot \nabla v^\alpha$, is treated as an independent nodal variable within the finite-element discretization.

## 3 Results

This section presents numerical simulations of single crystals incorporating the three slip systems specified in Table 1. Shear is applied at a constant strain rate of $2 \times 10^{-4}$.

The computational domain is discretized using plane-strain quadratic elements (eight-node elements with nine integration points), as shown in Fig. 1a. The selected mesh provides satisfactory insensitivity with respect to element size while maintaining computational efficiency. To impose microscopically hard boundary conditions, a surrounding layer with a thickness equal to 1% of the crystal dimension is introduced, enforcing zero plastic slip along the boundary.

Throughout the analysis, conventional isotropic hardening associated with slip resistance evolution is neglected, i.e., $\dot{S}^\alpha = 0$ in Eq. (18). The material and model parameters adopted in the simulations are summarized in Table 2. The influence of activating a passivation layer at a prescribed strain level is examined in Section 3.1, with representative results presented in Fig. 2.

## 3.1 Imposing a passivation layer at a specific strain

The proposed flow rule contains two gradient contributions: a recoverable energetic microstress and a dissipative microstress. To examine the influence of passivation, the crystal is first subjected to simple shear deformation (Fig. 1b) without any boundary layer and without constraints on plastic slip. After reaching a prescribed strain level, a surrounding microhard boundary layer is introduced, and the loading is continued. The resulting response is compared with that obtained when the microhard boundary layer is enforced from the onset of deformation.

The mechanical response for four representative parameter sets, labeled A through D, is shown in Fig. 2a. Enlarged views highlighting the onset of plastic yielding and the activation of passivation are provided in Figs. 2b and 2c, respectively. Data sets A and B correspond to cases in which passivation is present throughout the entire deformation process. In contrast, data sets C and D represent scenarios where passivation is imposed only after 2.5% shear strain.

For parameter choices $L_1 \neq 0$ and $L_2 = 0$ (data sets B and D), the model exhibits gradient hardening without gradient strengthening. As expected, no extension of the initial elastic regime is observed in these cases. When both length-scale parameters are active ($L_1 = L_2 \neq 0$, data sets A and C), the formulation additionally produces gradient strengthening.

A key observation is that the strengthening at initial yield (approximately 0.5% shear strain in data set A) coincides quantitatively with the elevated response observed at the instant of passivation activation (2.5% shear strain in data set C). The dependence on internal length scale is clearly reflected in these results. Moreover, beyond 2.5% strain, the stress–strain curves of data sets C and D closely reproduce the behavior of data sets A and B beyond 0.5% strain, indicating a shift in the effective response upon passivation.

Further insight is obtained by examining data points G, J, K, and L, marked in Fig. 2 and summarized in Table 3. Points G and K display nearly identical distributions and magnitudes of GND densities, despite K corresponding to a larger accumulated plastic deformation. A similar correspondence is observed between points J and L, even though L has undergone more extensive plastic flow. Notably, points K and L (at 4% strain) include deformation from 0% to 2.5% strain under non-passivated conditions, during which no boundary-induced GND accumulation occurred.

These results indicate that constitutive models capable of reproducing scale-dependent strengthening at the onset of plastic flow inherently generate a sharp, quasi-elastic response when passivation is activated. Conversely, formulations that do not predict gradient strengthening at initial yield are unable to capture the elevated boundary-induced response associated with passivation.

## 4  Summary and Conclusion

This study has presented a finite-deformation gradient crystal plasticity framework for single crystals, formulated within the theoretical structure introduced by Gurtin. The model was employed to investigate the mechanical response of crystals subjected to passivation-type boundary constraints.The analysis demonstrates that constitutive formulations capable of reproducing scale-dependent strengthening at the initiation of plastic flow naturally predict a pronounced increase in stiffness when passivation is activated during deformation. This establishes a direct connection between gradient-induced yield strengthening and boundary-driven elevation of the mechanical response.Although the observations obtained here suggest a general underlying mechanism, further assessment across alternative gradient crystal plasticity models would be valuable. Systematic benchmark comparisons could help determine the broader applicability of the proposed interpretation.

## Acknowledgment

This work was partially supported by the German Research Foundation (DFG) under grant PO-2242/3-1.

Table 1. Predefined slip systems, $\boldsymbol{s}^\alpha$ and $\boldsymbol{m}^\alpha$.

Table 2. A common set of material coefficients and modeling parameters.

Table 3. Data points G, J, K, and L, indicated in Fig. 2 are detailed (online color figure).

Fig. 1. Fig. 2.

# References

- Aifantis, E. C. (1984). On the microstructural origin of certain inelastic models. *Journal of Engineering Materials and Technology*, **106**, 326–330.
- Bonet, J., & Wood, R. D. (2008). *Nonlinear Continuum Mechanics for Finite Element Analysis* (2nd ed.). Cambridge University Press.
- Evans, A. G., & Hutchinson, J. W. (2009). A critical assessment of theories of strain gradient plasticity. *Acta Materialia*, **57**, 1675–1688.
- Fleck, N. A., & Hutchinson, J. W. (1993). A phenomenological theory for strain gradient effects in plasticity. *Journal of the Mechanics and Physics of Solids*, **41**, 1825–1857.
- Forest, S. (2008). Some links between Cosserat, strain gradient crystal plasticity and the statistical theory of dislocations. *Philosophical Magazine*, **88**, 3549–3563.
- Gurtin, M. E. (2006). The Burgers vector and the flow of screw and edge dislocations in finite-deformation single-crystal plasticity. *Journal of the Mechanics and Physics of Solids*, **54**, 1882–1898.
- Gurtin, M. E. (2008). A finite-deformation, gradient theory of single-crystal plasticity with free energy dependent on densities of geometrically necessary dislocations. *International Journal of Plasticity*, **24**, 702–725.
- Gurtin, M. E. (2010). A finite-deformation, gradient theory of single-crystal plasticity with free energy dependent on the accumulation of geometrically necessary dislocations. *International Journal of Plasticity*, **26**, 1073–1096.
- Gurtin, M. E., & Anand, L. (2009). Thermodynamics applied to gradient theories involving the accumulated plastic strain: The theories of Aifantis and Fleck and Hutchinson and their generalization. *Journal of the Mechanics and Physics of Solids*, **57**, 405–421.
- Gurtin, M. E., & Reddy, B. D. (2016). Some issues associated with the intermediate space in single-crystal plasticity. *Journal of the Mechanics and Physics of Solids*, **95**, 230–238.
- Huang, Y., Gao, H., Nix, W. D., & Hutchinson, J. W. (2000). Mechanism-based strain gradient plasticity—II. Analysis. *Journal of the Mechanics and Physics of Solids*, **48**, 99–128.
- Kuroda, M., & Tvergaard, V. (2008). A finite deformation theory of higher-order gradient crystal plasticity. *Journal of the Mechanics and Physics of Solids*, **56**, 2573–2584.

- Reddy, B. D., Ebobisse, F., & McBride, A. (2008). Well-posedness of a model of strain gradient plasticity for plastically irrotational materials. *International Journal of Plasticity*, **24**, 55–73.
- Voyiadjis, G. Z., & Song, Y. (2019). Strain gradient continuum plasticity theories: Theoretical, numerical and experimental investigations. *International Journal of Plasticity*, **121**, 21–59.

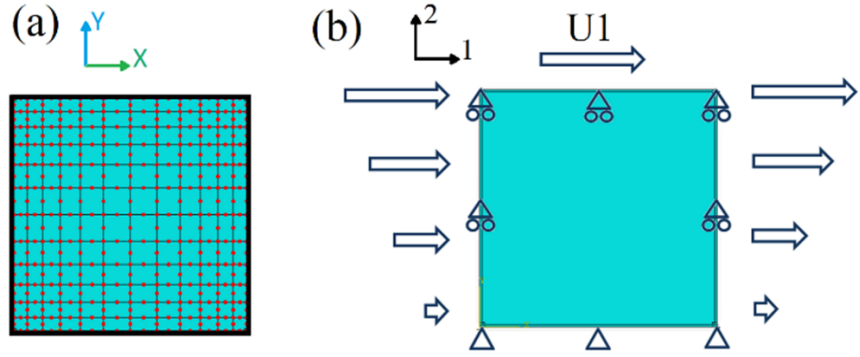

Fig. 1. (a) Finite-element model of the single crystal surrounded by a microhard boundary layer (black outline). Red dots indicate nodal points. Each quadratic element contains nine integration points. (b) Schematic representation of the applied simple shear deformation mode.

Table 1. Predefined slip systems, $\mathbf{s}^\alpha$ and $\mathbf{m}^\alpha$.

| Slip-system number $\alpha$ | Slip vector $\mathbf{s}^\alpha$ | Normal slip plane vector $\mathbf{m}^\alpha$ |
|---|---|---|
| 1 | $(\sqrt{3}/2, 1/2)$ | $(-1/2, \sqrt{3}/2)$ |
| 2 | $(1/2, \sqrt{3}/2)$ | $(\sqrt{3}/2, -1/2)$ |
| 3 | $(-1/2, \sqrt{3}/2)$ | $(\sqrt{3}/2, 1/2)$ |

Table 2. A common set of material coefficients and modeling parameters.

| $\mu$ | $\lambda$ | $S_0$ | $S^\alpha, S_y$ | $\dot{S}^\alpha$ | $S_d$ | $\mathbb{M}$ | $m$ | $q$, n | $v_0$ | $d_0$ |
|---|---|---|---|---|---|---|---|---|---|---|
| 76.9 GPa | 115.3 GPa | 0.9 MPa | 200 MPa | 0 MPa/s | 0.2 MPa | 0.2 MPa | 0.05 - | 1.0 - | 1e-3 $s^{-1}$ | 5e-4 $s^{-1}$ |

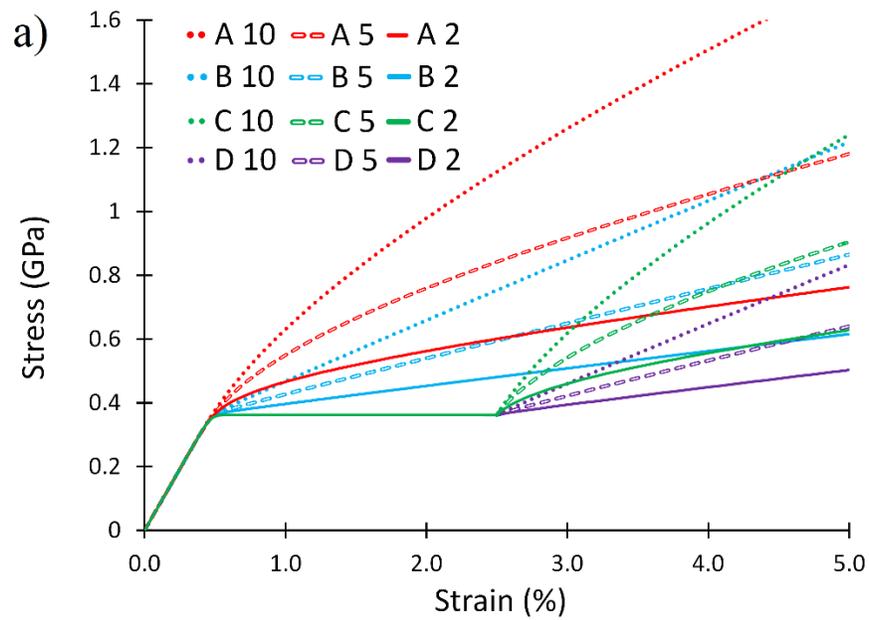

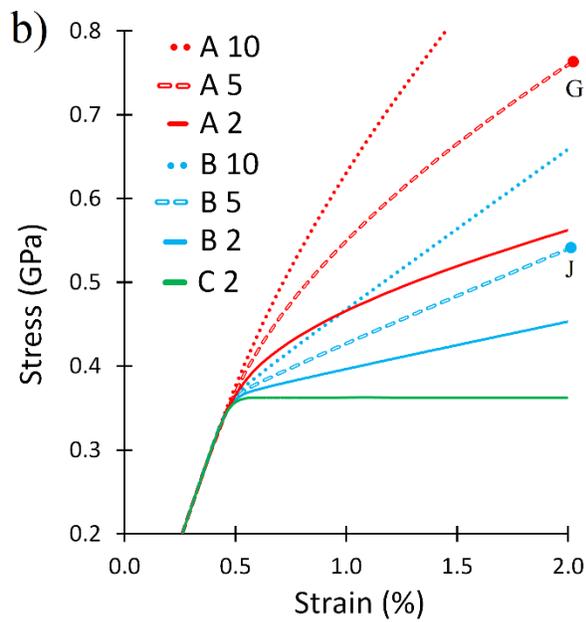
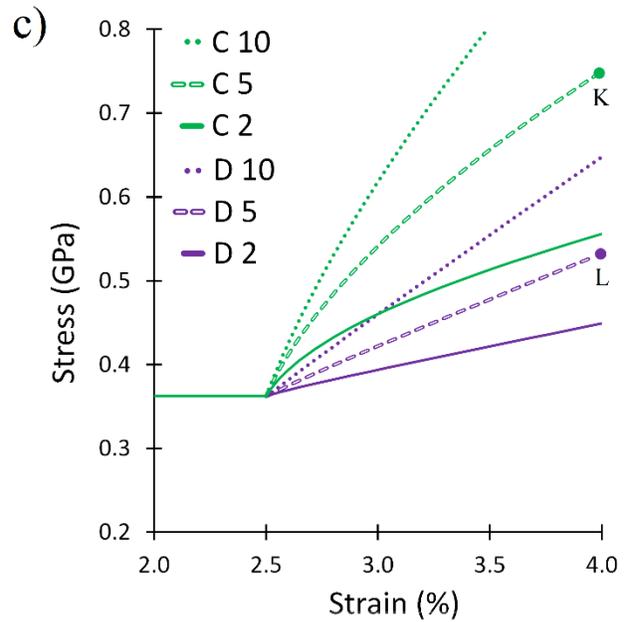

| Passivation from the beginning of deformation | | | Passivation at the shear strain of 2.5% | | |
|---|---|---|---|---|---|
| Data set | $L_1$ ($\mu m$) | $L_2$ ($\mu m$) | Data set | $L_1$ ($\mu m$) | $L_2$ ($\mu m$) |
| A10 | 10 | 10 | C10 | 10 | 10 |
| A 5 | 5 | 5 | C 5 | 5 | 5 |
| A 2 | 2 | 2 | C 2 | 2 | 2 |
| B10 | 10 | 0 | D10 | 10 | 0 |
| B 5 | 5 | 0 | D 5 | 5 | 0 |
| B 2 | 2 | 0 | D 2 | 2 | 0 |

Fig. 2. Stress–strain response of the single crystal subjected to the simple shear deformation illustrated in Fig. 1b. Four parameter sets (A–D) are considered. (a) presents the complete responses, while (b) and (c) provide enlarged views highlighting the onset of plastic yielding and the activation of passivation, respectively. Data sets A and B correspond to cases where the passivation boundary condition is applied from the beginning of deformation. Data sets C and D represent cases in which passivation is introduced after 2.5% shear strain. When $L_2 \neq 0$ (data sets A and C), gradient strengthening is present. For $L_2 = 0$ (data sets B and D), only gradient hardening is observed. The post-passivation response of data sets C and D beyond 2.5% strain closely reproduces the behavior of data sets A and B beyond 0.5% strain. Data points G, J, K, and L are identified in the figure and further quantified in Table 3 (online color version).

Table 3. Data points G, J, K, and L, indicated in Fig. 2 are detailed (online color figure).

| Data point | G | K | J | L |
|---|---|---|---|---|
| Data set | A 5 | C 5 | B 5 | D 5 |
| Strain % | 2 | 4 | 2 | 4 |
| Accumulated GND Density ($mm^{-1}$) | 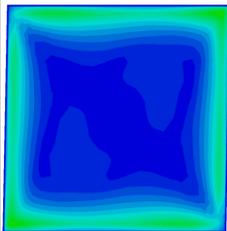 | 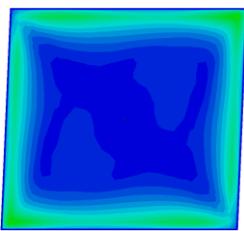 | 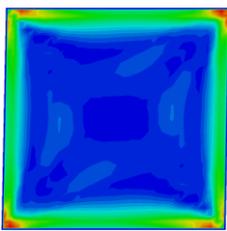 | 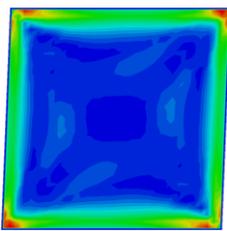 |
| Directional plastic flow, slip system 1 in Table 1 | 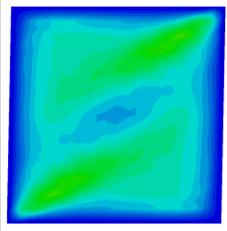 | 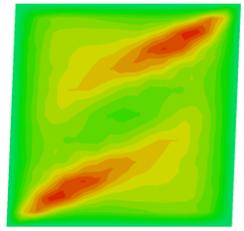 | 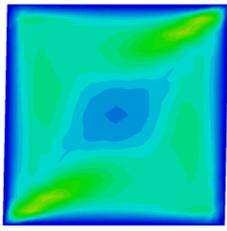 | 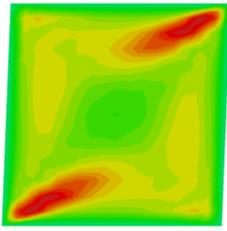 |